\documentclass[prd,aps,showpacs,notitlepage,superscriptaddress,nofootinbib]{revtex4-1}
\usepackage{epsfig}
\usepackage{color}
\usepackage{url}
\usepackage{amsmath}
\usepackage{multirow}

\def\bea#1\eea{\begin{align}#1\end{align}}

\makeatletter
\def\slash#1{{\mathpalette\c@ncel{#1}}} % TeXbook, bottom of p360
\makeatother

\newcommand\beq{\begin{eqnarray}}
\newcommand\eeq{\end{eqnarray}}

\newcommand\la{\langle}
\newcommand\ra{\rangle}

\begin{document}
\title{Introduction to the transverse-momentum-weighted technique in the twist-3 collinear factorization approach}

\date{\today}
                       
\author{Hongxi Xing}
\email{hxing@m.scnu.edu.edu}
\affiliation{Institute of Quantum Matter and School of Physics and Telecommunication Engineering,
South China Normal University, Guangzhou 510006, China} 

\author{Shinsuke Yoshida}
\email{shinyoshida85@gmail.com}
\affiliation{Institute of Quantum Matter and School of Physics and Telecommunication Engineering,
South China Normal University, Guangzhou 510006, China}

\begin{abstract}
The twist-3 collinear factorization framework has drawn much attention in recent decades as a successful approach in describing the data for single spin asymmetries (SSAs). Many SSAs data have been experimentally accumulated in a variety of energies since the first measurement was done in late 70s and it is expected that the future experiments like Electron-Ion collider will provide us with more data. In order to perform a consistent and precise description of the data taken in different kinematic regimes, the scale evolution of the collinear twist-3 functions and the perturbative higher order hard part coefficients are mandatory. In this paper, we introduce the techniques for next-to-leading order (NLO) calculation of transverse-momentum-weighted SSAs, which can be served as a useful tool to derive the QCD evolution equation for twist-3 functions, and to verify the QCD collinear factorization for twist-3 observables at NLO, as well as to obtain the finite NLO hard part coefficients.   
\end{abstract}

\maketitle

\section{Introduction}

%Single transverse spin asymmetries (SSAs) has attracted tremendous attention due to its sensitivity to probe the proton's three-dimensional fundamental structures, as the transverse spin can correlate with the transverse momentum of the partons inside a polarized proton. 
The large Single transverse spin asymmetries (SSAs) have been a longstanding problem over 40 years since it was turned out that the conventional perturbative calculation based on the parton model picture failed to describe the large SSAs which were experimentally observed in pion and polarized hyperon productions ~\cite{Klem:1976ui,Bunce:1976yb}. In the recent several years, two QCD factorization frameworks have been proposed to study phenomenologically the observed SSAs: the transverse momentum dependent factorization approach \cite{TMD-fac,Brodsky,MulTanBoe,Boer:1997nt, Anselmino:2009st} and the twist-3 collinear factorization approach \cite{Efremov,qiu,Kanazawa:2000cx,koike,Koike:2011mb,Yuan:2009dw,Liang:2012rb}. These two frameworks are shown to be equivalent in the common applied kinematic region \cite{unify}.

The twist-3 collinear factorization framework is a natural extension of the conventional 
perturbative QCD framework and it could give a reasonable description of the large SSAs. 
Measurements of SSAs at Relativistic-Heavy-Ion-Collider(RHIC)~\cite{Adams:2003fx,Adler:2005in,Arsene:2008aa} 
have greatly motivated the theoretical work on developing the twist-3 framework, 
because it is a unique applicable framework for single hadron productions in proton-proton 
collision. A series of important work have been done in the past a few decades
and the SSAs for the hadron production was completed at leading-order (LO) with respect to the QCD 
strong coupling constant $\alpha_s$~\cite{Efremov,qiu,Kanazawa:2000cx,koike,Koike:2011mb,Yuan:2009dw,Liang:2012rb} . Recent numerical simulations based on the complete LO result confirmed that the twist-3 approach gives a reasonable description of the SSA data provided by RHIC~\cite{Kanazawa:2014dca,Gamberg:2017gle}. 

Electron-Ion-Collider (EIC) is a next-generation hadron collider expected to provide more data 
in different kinematic regimes for SSAs. In order to extract the fundamental structure of the 
nucleon from the measurements at a future EIC, comprehensive and precise calculations for SSAs 
in transversely polarized lepton-proton collision is highly demanded. It's well known that 
nonperturbative functions in the perturbative QCD calculation, in general, receive logarithmic 
radiative corrections and the evolution equation with respect to this logarithmic scale are 
necessary for a systematic treatment of the cross sections in wide range of energies. Most 
famous example is the DGLAP evolution equation of the twist-2 parton distribution functions 
(PDFs). Correct description of the small scale violation of the structure function controlled 
by the DGLAP equation was an important success of the QCD phenomenology in the early days. The 
twist-3 function is expected to have similar logarithmic dependence and its evolution equation 
will play an important role in global fitting of the SSA data accumulated in different energies. 
Consistent description of the data will be a good evidence that the twist-3 framework, 
one of major fundamental developments in  recent QCD phenomenology, is a feasible theory to 
solve the 40-years mystery in high energy physics. The evolution equations for the twist-3 
functions have been derived in two different methods. The first method is a calculation of the 
higher-order corrections to the nonperturbative function itself \cite{Kang:2008ey,Zhou:2008mz,Braun:2009mi,Schafer:2012ra,Ma:2012xn,Kang:2012em,Kang:2010xv,Ma:2017upj}. 
Since the nonperturbative function has the operator definition, we can investigate the infrared 
singularity of the operator through higher-order perturbative calculation. We can read the 
evolution equation from the infrared structure of the function. This is a standard technique 
and the evolution equations have been derived for the twist-3 distribution functions for initial 
state proton \cite{Kang:2008ey,Zhou:2008mz,Braun:2009mi,Schafer:2012ra,Ma:2012xn,Kang:2012em} and the twist-3 fragmentation functions for final state hadron 
\cite{Kang:2010xv,Ma:2017upj}. The second method which we will review in this paper is a 
transverse-momentum-weighted technique for the SSAs \cite{Vogelsang:2009pj,Kang:2012ns,Yoshida:2016tfh,Dai:2014ala,Chen:2016dnp,Chen:2017lvx}. 
Except for deriving the QCD evolution equation for twist-3 nonperturbative functions, 
the transverse-momentum-weighted technique can be also used as a tool to verify the twist-3 
collinear factorization at higher orders in strong coupling constant $\alpha_s$. 
There is also phenomenological interest related to this technique. One can use the standard 
dimensional regularization method to derive the NLO hard part coefficient for transverse 
momentum weighted SSAs, which can be used for high precision extraction of twist-3 
functions from the relevant experimental data. The recent measurement of the 
transverse-momentum-weighted SSAs at COMPASS ~\cite{Alexeev:2018zvl} strongly motivates the phenomenological 
application of the results reviewed in this paper. We expect more data will be produced 
in future COMPASS and EIC measurements.

The rest of the paper is organized as follows. In Sec. \ref{sec-LO} we present the notation and the calculation of transverse-momentum-weighted SSAs at leading order for semi-inclusive deep inelastic scattering (SIDIS). In Sec. \ref{sec-nlo} we present the detail of NLO calculation for both real and virtual corrections, we show the cancellation of soft divergence in the sum of real and virtual corrections, and the collinear divergences can be absorbed into the redefinition of twist-3 Qiu-Sterman function and unpolarized leading twist fragmentation function. In Sec. \ref{sec-app} we review the application of the transverse-momentum-weighted technique to other processes that have been done in recent years. We conclude our paper in Sec. \ref{sec-sum}.

%--------------------- Section 2 ----------------

\section{Transverse-momentum-weighted SSA at leading order}
\label{sec-LO}
In this paper, we take the process of SIDIS as an example to show the techniques of perturbative calculation for transverse-momentum-weighted differential cross section at twist-3. We start this section by specifying our notation and the kinematics of SIDIS, and present the calculation for transverse-momentum-weighted SSA at leading order (LO).

\subsection{Notation}
 We consider the scattering of an unpolarized lepton with momentum $l$ on a transversely polarized proton with momentum $p$ and transverse spin $S_{\perp}$, and observe the final state hadron production with momentum $P_h$,
\beq
e(l)+p^{\uparrow}(p,S_{\perp})\to e(l')+h(P_h)+X.
\eeq
We focus on one-photon exchange process with the momentum of the virtual photon given by $q=l-l'$ and its invariant mass $Q^2=-q^2$. We define all vectors in the so-called hadron frame. 
%It's convenient to choose the hadron frame where all vectors are assigned as
%\beq
%l&=&{Q\over 2}(\cosh\psi,\sinh\psi\cos\phi,\sinh\psi\sin\phi,-1),
%\\
%q&=&(0,0,0,-Q),\hspace{5mm}p^{\mu}=\Bigl({Q\over 2x_B},0,0,{Q\over 2x_B}\Bigr),
%\hspace{5mm}S^{\mu}_{\perp}=(0,\cos\Phi_S,\sin\Phi_S,0),
%\\
%P_h&=&{z_hQ\over 2}\Bigl(1+{P^2_{h\perp}\over z_h^2Q^2}
%,{2P_{h\perp}\over z_hQ}\cos\chi,{2P_{h\perp}\over z_hQ}\sin\chi,{P^2_{h\perp}\over z_h^2Q^2}-1\Bigr),
%\eeq
%where $\cosh\psi={2x_BS_{ep}\over Q^2}-1$, $\phi$ is the azimuthal angle for the outgoing lepton, $\chi$ and $\Phi_S$ are the azimuthal angles for the final state hadron momentum $P_h$ and spin vector $S_{\perp}$. 
We define $p_c=P_h/z$ to be the momentum for the parton that fragments into the final state hadron. The conventional Lorentz invariant variables in SIDIS are defined as
\beq
S_{ep}=(p+l)^2,\hspace{5mm}x_B={Q^2\over 2p\cdot q},\hspace{5mm}
z_h={p\cdot P_h\over p\cdot q},\hspace{5mm}y=\frac{p\cdot q}{p\cdot l}.
\eeq
For clear understanding, we start with the $P_{h\perp}$-integrated cross section at leading twist in unpolarized lepton-proton scattering
\beq
{d\sigma\over dx_Bdydz_h}
=\int d^2P_{h\perp}{d\sigma\over dx_Bdydz_hd^2P_{h\perp}}.
\eeq
There is only one hard scale $Q^2$ in this case, therefore the differential cross section shown above can be reliably computed by using the standard collinear factorization formalism. The LO contribution is given by $2\to 1$ scattering amplitude $\gamma^*+q\to q$.
%%%%%%%%%%%%%%%%%%%%%%%%%%%%%%%%%%%%%%%%%%%%%%%%%%%%%%%%%%%%%%%%%%%%%%%%%%%%%%%%%%%%%%%%%%%%%%%%%
%\begin{figure}[h]
%\begin{center}
%  \includegraphics[height=4cm,width=6cm]{fig1.eps}\hspace{1cm}
%\end{center}
 %\caption{Leading order diagram for the $P_{h\perp}$-weighted cross section.}
%\end{figure}
%%%%%%%%%%%%%%%%%%%%%%%%%%%%%%%%%%%%%%%%%%%%%%%%%%%%%%%%%%%%%%%%%%%%%%%%%%%%%%%%%%%%%%%%%%%%%%%%% 
It is trivial that the LO cross section is proportional to the unpolarized PDFs and the unpolarized fragmentation
functions (FFs),
\bea
\frac{d\sigma^{LO}}{dx_Bdydz_h} = \sigma_0\sum_q f_{q/p}(x_B,\mu^2)D_{q\to h}(z_h,\mu^2),
\eea
where $\sigma_0$ is the LO Born cross section $\sigma_0=\frac{2\pi\alpha_{em}^2}{Q^2}\frac{1+(1-y)^2}{y}$ with $\alpha_{em}= {e^2\over 4\pi}$ is the QED coupling constant. The bare results at $\mathcal{O}(\alpha_s)$ contain infrared divergences which represent the long-range interaction in hadronic collision process. These divergences are canceled by the renormalization of the PDFs and FFs and the DGLAP evolution equations are derived as the renormalization group equations. The final result at NLO can be written as the convolution of finite hard part coefficient $H$ and nonperturbative functions (PDFs and FFs) \cite{Kang:2012ns}
\bea
\frac{d\sigma^{NLO}}{dx_Bdydz_h} = \sum_{i,j} f_{i/p}\otimes H_{\gamma^*+i\to j+k}^{NLO}\otimes D_{j\to h}.
\eea
where $\otimes$ represents for convolution.

The concept of the transverse-momentum-weighted technique is mostly the same with the twist-2 case. Notice that direct $P_{h\perp}$-integration of the cross section for unpolarized lepton scattering off transversely polarized proton vanishes due to the linear dependence of $P_h$. Realize that the SSA is characterized in terms of three vectors: the momentum of the final state hadron, the momentum and the spin of the initial state proton, which can be combined as
\beq
\epsilon^{\alpha\beta\rho\sigma}P_{h\alpha}S_{\perp\beta}\,p_{\rho}n_{\sigma}
=\epsilon^{ij}P_{h\perp i}S_{\perp j}
\equiv \epsilon^{P_{h\perp}S_{\perp}pn},
\eeq
where $\epsilon^{ij}$ is a two-dimensional antisymmetric tensor with $\epsilon^{12}=1$, $n$ is an arbitrary vactor satisfies $p\cdot n=1$ and $n^2=0$. 
We introduce a weight factor $\epsilon^{P_{h\perp}S_{\perp}pn}$ and consider the following transverse momentum weighted differential cross section
\beq
{d\la P_{h\perp}\Delta\sigma\ra\over dx_Bdydz_h}
\equiv\int d^2P_{h\perp}\epsilon^{P_{h\perp}S_{\perp}pn}
{d\Delta\sigma\over dx_Bdydz_hd^2P_{h\perp}},
\eeq
which is well defined after $P_{h\perp}$-integration.
Since the virtuality $Q^2$ is the only hard scale after the $P_{h\perp}$-integration, one can safely use the collinear twist-3 factorization formalism, and the technique in performing NLO calculation will follow those used at leading twist. Same technique has been applied to Drell-Yan dilepton production in proton-proton collisions \cite{Vogelsang:2009pj,Chen:2016dnp} and can be extended to polarized electron-positron collisions.

We recall the cross section for SIDIS presented in~\cite{Yoshida:2016tfh},
\beq
{d\Delta\sigma\over dx_Bdydz_hd^2P_{h\perp}}
&=&{\alpha^2_{em}\over 128\pi^4z_hx_B^2S^2_{ep}Q^2}L_{\mu\nu}W^{\mu\nu},
\eeq
where $L_{\mu\nu}=2(l_{\mu}l'_{\nu}+l_{\nu}l'_{\mu})-Q^2g_{\mu\nu}$ is the leptonic tensor. We focus on the metric contribution $L_{\mu\nu}\to -Q^2 g_{\mu\nu}$ and the SSA generated by initial state twist-3 distribution functions of the transversely polarized proton. Then we can factorize the nonperturbative part by introducing the usual twist-2 unpolarized fragmentation function
\beq
W^{\mu\nu}&=&\sum_i\int{{dz\over z^2}}w_i^{\mu\nu} D_{i\to h}(z).
\eeq
The hadronic tensor $w_i^{\mu\nu}$ describes a scattering of the virtual photon and the transversely polarized proton. We will make the subscript $i$ implicit in the rest part of this paper for simplicity.

\subsection{Leading order}
We demonstrate how to derive the LO cross section for the transverse-momentum-weighted SSA based on the collinear twist-3 framework and show that the LO cross section is proportional to the first moment of the TMD Sivers function.
The twist-3 calculation is well formulated in the diagramatic method. 
%%%%%%%%%%%%%%%%%%%%%%%%%%%%%%%%%%%%%%%%%%%%%%%%%%%%%%%%%%%%%%%%%%%%%%%%%%%%%%%%%%%%%%%%%%%%%%%%%
\begin{figure}[h]
\begin{center}
  \includegraphics[height=4cm,width=12cm]{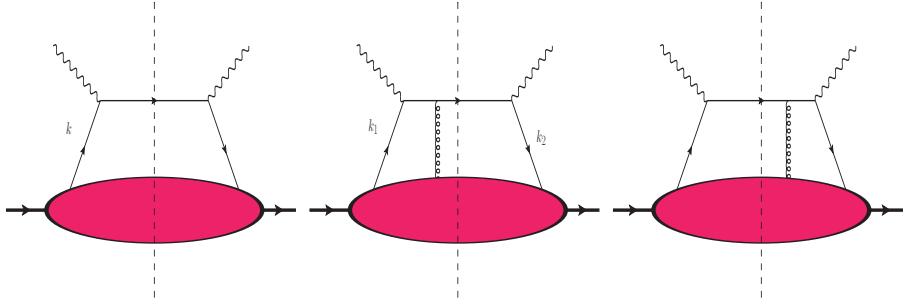}\hspace{1cm}
\end{center}
 \caption{A series of LO diagrams in the diagrammatic method.}
 \label{fig-LO}
\end{figure}
%%%%%%%%%%%%%%%%%%%%%%%%%%%%%%%%%%%%%%%%%%%%%%%%%%%%%%%%%%%%%%%%%%%%%%%%%%%%%%%%%%%%%%%%%%%%%%%%% 
We consider a set of the general diagrams shown in Fig. \ref{fig-LO} and extract twist-3 contributions from these diagrams. We start from the first diagram in Fig. \ref{fig-LO},  which can be expressed as 
\beq
w^{\mu\nu}_1=\int d^4\xi\int{d^4k\over (2\pi)^4}e^{ik\cdot\xi}
\la pS_{\perp}|\bar{\psi}_j(0)\psi_i(\xi)|pS_{\perp}\ra 
H^{\mu\nu}_{ji}(k)\delta^2\left(k_{\perp}-p_{c\perp}\right),
\label{first}
\eeq
where $k$ and $p_c$ are the momenta of the parton from initial state proton and that fragments to the final state observed hadron, respectively, $p_{c\perp}=P_{h\perp}/z$. The hard part at LO is given by
\beq
H^{\mu\nu}_{ji}(k)=\left[\gamma^{\nu}(\slash{k}+\slash{q})\gamma^{\mu}\right]_{ji}
(2\pi)^4\delta\left(k^++q^+-p_c^+\right)\delta(k^-+q^--p_c^-).
\eeq
We perform $P_{h\perp}$-integration before the collinear expansion,
\beq
\int d^2P_{h\perp}\epsilon^{P_{h\perp}S_{\perp}pn}
\delta^2(k_{\perp}-p_{c\perp})=z^3\epsilon^{\alpha S_{\perp}pn}k_{\perp\alpha}.
\eeq
Because the $P_{h\perp}$-integration gives $O(k_{\perp})$ factor, we can identify the leading term in the 
collinear expansion as a twist-3 contribution. We perform the collinear expansion of the hard part around $k^{\mu}=(k\cdot n)p^{\mu} \equiv k_p^{\mu}$
\beq
H^{\mu\nu}(k)=H^{\mu\nu}(k_p)+O(k_{\perp}),
\eeq
and substitute the above expansion into the hadronic tensor as shown in Eq. (\ref{first})
\beq
\int d^2P_{h\perp}\epsilon^{P_{h\perp}S_{\perp}pn}w_1^{\mu\nu}
&=&z^3\epsilon_{\alpha}^{\ S_{\perp}pn}
\int d^4\xi\int{d^4k\over (2\pi)^4}e^{ik\cdot\xi}
\la pS_{\perp}|\bar{\psi}_j(0)\psi_i(\xi)|pS_{\perp}\ra 
k_{\perp}^{\alpha}H^{\mu\nu}_{ji}(k_p)
\nonumber\\
&=&iz^3\epsilon_{\alpha}^{\ S_{\perp}pn}\int dx\int{d\lambda\over 2\pi} e^{i\lambda x}
\la pS_{\perp}|\bar{\psi}_j(0)\partial^{\alpha}\psi_i(\lambda)|pS_{\perp}\ra 
H^{\mu\nu}_{ji}(xp).
\label{first_result}
\eeq
Now we turn to the second and the third diagrams in Fig. \ref{fig-LO}. These two diagrams can be expressed as
\beq
w_2^{\mu\nu}&=&\int d^4\xi_1\int d^4\xi_2\int{d^4k_1\over (2\pi)^4}\int{d^4k_2\over (2\pi)^4}
e^{ik_1\cdot\xi_1}e^{i(k_2-k_1)\cdot\xi_2}
\la pS_{\perp}|\bar{\psi}_j(0)gA^{\rho}(\xi_2)\psi_i(\xi_1)|pS_{\perp}\ra
\nonumber\\
&&\times\left[H^{\mu\nu}_{L\rho\,ji}(k_1,k_2)\delta^2(k_{2\perp}-p_{c\perp})
+H^{\mu\nu}_{R\rho\,ji}(k_1,k_2)\delta^2(k_{1\perp}-p_{c\perp})\right],
\label{second}
\eeq
where the hard parts are given by
\beq
H^{\mu\nu}_{L\rho\,ji}(k_1,k_2)&=&-\left[\gamma^{\nu}(\slash{k}_2+\slash{q})\gamma_{\rho}
(\slash{k}_1+\slash{q})\gamma^{\mu}\right]_{ji}
{1\over (q+k_1)^2+i\epsilon}
(2\pi)^4\delta(k_2^++q^+-p_c^+)\delta(k_2^-+q^--p_c^-),
\nonumber\\
H^{\mu\nu}_{R\rho\,ji}(k_1,k_2)&=&-\left[\gamma^{\nu}(\slash{k}_2+\slash{q})\gamma_{\rho}
(\slash{k}_1+\slash{q})\gamma^{\mu}\right]_{ji}
{1\over (q+k_2)^2-i\epsilon}
(2\pi)^4\delta(k_1^++q^+-p_c^+)\delta(k_1^-+q^--p_c^-).
\eeq
The $P_{h\perp}$-integration gives $O(k_{1,2\perp})$ and then the leading term in collinear
expansion gives twist-3 contribution again,
\beq
H^{\mu\nu}_{L(R)\rho\,ji}(k_1,k_2)
=H^{\mu\nu}_{L(R)\rho\,ji}(k_{1p},k_{2p})+O(k_{1,2\perp}).
\eeq
For the matrix element, we have to separate the components of the gluon field $A^{\rho}$ into ``longitudinal" and ``transverse" part as
\beq
A^{\rho}=A^np^{\rho}+(A^{\rho}-A^np^{\rho}).
\eeq 
The longitudinal part $A^np^{\rho}$ gives the leading contribution. It is straightforward to derive 
the Ward-Takahashi identities(WTIs) for the hard parts,
\beq
p^{\rho}H^{\mu\nu}_{L\rho\,ji}(k_1,k_2)&=&[\gamma^{\nu}(x_2\slash{p}+\slash{q})\gamma^{\mu}]_{ji}
{1\over x_2-x_1-i\epsilon}(2\pi)^4{2x_B\over Q^2}\delta(x_2-x_B)\delta(1-\hat{z})
\nonumber\\
&=&{1\over x_2-x_1-i\epsilon}H^{\mu\nu}_{ji}(x_2p),
\nonumber\\
p^{\rho}H^{\mu\nu}_{R\rho\,ji}(k_1,k_2)&=&-[\gamma^{\nu}(x_1\slash{p}+\slash{q})\gamma^{\mu}]_{ji}
{1\over x_2-x_1-i\epsilon}(2\pi)^4{2x_B\over Q^2}\delta(x_1-x_B)\delta(1-\hat{z})
\nonumber\\
&=&-{1\over x_2-x_1-i\epsilon}H^{\mu\nu}_{ji}(x_1p).
\label{WTI}
\eeq
Finally the hadronic tensor shown in Eq. (\ref{second}) can be expressed as
\beq
&&\int d^2P_{h\perp}\epsilon^{P_{h\perp}S_{\perp}pn} w_2^{\mu\nu}
\nonumber\\
&=&z^3\epsilon_{\alpha}^{\ S_{\perp}pn}\int d^4\xi_1\int d^4\xi_2\int{d^4k_1\over (2\pi)^4}\int{d^4k_2\over (2\pi)^4}
e^{ik_1\cdot\xi_1}e^{i(k_2-k_1)\cdot\xi_2}
\la pS_{\perp}|\bar{\psi}_j(0)gA^n(\xi_2)\psi_i(\xi_1)|pS_{\perp}\ra
\nonumber\\
&&\times{1\over x_2-x_1-i\epsilon}\Bigl[k^{\alpha}_{2\perp}H^{\mu\nu}_{ji}(x_2p)
-k^{\alpha}_{1\perp}H^{\mu\nu}_{ji}(x_1p)
\Bigr]
\nonumber\\
&=&z^3\epsilon_{\alpha}^{\ S_{\perp}pn}\int d^4\xi_1\int d^4\xi_2\int{d^4k_1\over (2\pi)^4}\int{d^4k_2\over (2\pi)^4}
e^{ik_1\cdot\xi_1}e^{i(k_2-k_1)\cdot\xi_2}
\la pS_{\perp}|\bar{\psi}_j(0)gA^n(\xi_2)\psi_i(\xi_1)|pS_{\perp}\ra
\nonumber\\
&&\times{1\over x_2-x_1-i\epsilon}\Bigl[(k^{\alpha}_{2\perp}-k^{\alpha}_{1\perp})H^{\mu\nu}_{ji}(x_2p)
+k^{\alpha}_{1\perp}(H^{\mu\nu}_{ji}(x_2p)-H^{\mu\nu}_{ji}(x_1p))\Bigr]
\nonumber\\
&=&iz^3\epsilon_{\alpha}^{\ S_{\perp}pn}\int dx\int{d\lambda\over 2\pi}
e^{i\lambda x}\la pS_{\perp}|\bar{\psi}_j(0)ig\int^{\infty}_{\lambda} d\lambda' 
\Bigl[\partial^{\alpha}A^n(\lambda' n)-\partial^nA^{\alpha}(\lambda' n)\Bigr]\psi_i(\lambda n)|pS_{\perp}\ra 
H^{\mu\nu}_{ji}(xp)
\nonumber\\
&&+iz^3\epsilon_{\alpha}^{\ S_{\perp}pn}\int dx\int{d\lambda\over 2\pi}
e^{i\lambda x}\la pS_{\perp}|\bar{\psi}_j(0)\Bigl[ig\int^{0}_{\lambda} d\lambda' 
A^n(\lambda' n)\Bigr]\partial^{\alpha}\psi_i(\lambda n)|pS_{\perp}\ra H^{\mu\nu}_{ji}(xp)
\nonumber\\
&&-iz^3\epsilon_{\alpha}^{\ S_{\perp}pn}\int dx\int{d\lambda\over 2\pi}
e^{i\lambda x}\la pS_{\perp}|\bar{\psi}_j(0)igA^{\alpha}(\lambda n)
\psi_i(\lambda n)|pS_{\perp}\ra H^{\mu\nu}_{ji}(xp).
\label{second_result}
\eeq
Combing Eqs. (\ref{first_result}) and (\ref{second_result}), we can obtain the result
\beq
&&\int d^2P_{h\perp}\epsilon^{P_{h\perp}S_{\perp}pn}w^{\mu\nu}
\nonumber\\
=&&iz^3\epsilon_{\alpha}^{\ S_{\perp}pn}\int dx\int{d\lambda\over 2\pi}
e^{i\lambda x}\la pS_{\perp}|\bar{\psi}_j(0)\Bigl(D^{\alpha}(\lambda n)
\psi_i(\lambda n)+ig\int^{0}_{\lambda} d\lambda' 
A^n(\lambda' n)\partial^{\alpha}\psi_i(\lambda n)
\Bigl)|pS_{\perp}\ra  H^{\mu\nu}_{ji}(xp)
\nonumber\\
&&+iz^3\epsilon_{\alpha}^{\ S_{\perp}pn}\int dx\int{d\lambda\over 2\pi}
e^{i\lambda x}\la pS_{\perp}|\bar{\psi}_j(0)ig\int^{\infty}_{\lambda} d\lambda'
\Bigl(\partial^{\alpha}A^n(\lambda' n)-\partial^nA^{\alpha}(\lambda' n)\Bigr)\psi_i(\lambda n)|pS_{\perp}\ra 
 H^{\mu\nu}_{ji}(xp).
\label{combined}
\eeq
We can find that this matrix element corresponds to $O(g)$ term in the first moment of the 
TMD correlator
\beq
&&\int d^2p_T\,p_T^{\alpha}\Bigl(\int{d\lambda\over 2\pi}\int{dx_T\over 2\pi}e^{i\lambda x}e^{ix_T\cdot p_T}
\la pS_{\perp}|\bar{\psi}_j(0)[0,\infty n][\infty n,\infty n+x_T][\infty n+x_T,\lambda n+x_T]
\psi_i(\lambda n+x_T)|pS_{\perp}\ra
\Bigr)
\nonumber\\
&=&i\int{d\lambda\over 2\pi}e^{i\lambda x}
\la pS_{\perp}|\bar{\psi}_j(0)[0,\lambda n]D^{\alpha}(\lambda n)\psi_i(\lambda n)|pS_{\perp}\ra
\nonumber\\
&&+i\int{d\lambda\over 2\pi}e^{i\lambda x}\int^{\infty}_{\lambda}
d\lambda'
\la pS_{\perp}|\bar{\psi}_j(0)[0,\lambda' n]igF^{\alpha n}(\lambda' n)[\lambda' n,\lambda n]\psi_i(\lambda n)|pS_{\perp}\ra
\nonumber\\
&=&-\pi{M_N\over 4}\epsilon^{\alpha pnS_{\perp}}G_{q,F}(x,x)+\cdots,
\eeq
where $[\cdots]$ represents the Wilson line, $M_N$ is the nucleon mass and 
we used the fact that the first moment of the TMD Sivers function gives the Qiu-Sterman function
$G_{q,F}(x,x)$. The nonlinear term in the field strength tensor 
$F^{\alpha +}$ and the higher order terms in the Wilson lines which have to be added to 
Eq. (\ref{combined}) come from the more gluon-linked diagrams in Fig. 2.
Finally we can derive the LO cross section formula as
\beq
{d\la P_{h\perp}\Delta\sigma\ra^{\rm LO}\over dx_Bdydz_h}
= \tilde\sigma_0\sum_qe^2_qG_{q,F}(x_B,x_B)D_{q\to h}(z_h).
\label{LOcross}
\eeq
where $\tilde\sigma_0 = -\pi M_N \sigma_0$. As we demonstrated here, the LO cross section for the 
weighted SSA is proportional to the first moment of the TMD function(also known as the kinematical
twist-3 function). We can expect the NLO 
contribution gives the evolution equation to the twist-3 Qiu-Sterman function, $G_{q,F}(x,x)$ in this case. This technique is 
quite general so that we can apply the same technique to other TMD functions.
When we focus on the twist-3 fragmentation effect, we can derive the evolution equation for the first Moment of the Collins function.
When we consider the double spin asymmetry $A_{LT}$ and change the weight factor 
$\epsilon^{ij}P_{h\perp i}S_{\perp j}\to (P_h\cdot S_{\perp})$, we can investigate other TMD
distribution functions like the Worm-Gear and the Pretzelosity.

%-----------------------   Section 3 ----------------------
\section{Transverse-momentum-weighted SSA at NLO}
\label{sec-nlo}
In this section, we review the calculation for transverse-momentum-weighted SSA at NLO including both real and virtual corrections. 

\subsection{Virtual correction}
\begin{figure}[h]
\begin{center}
  \includegraphics[height=4cm,width=12cm]{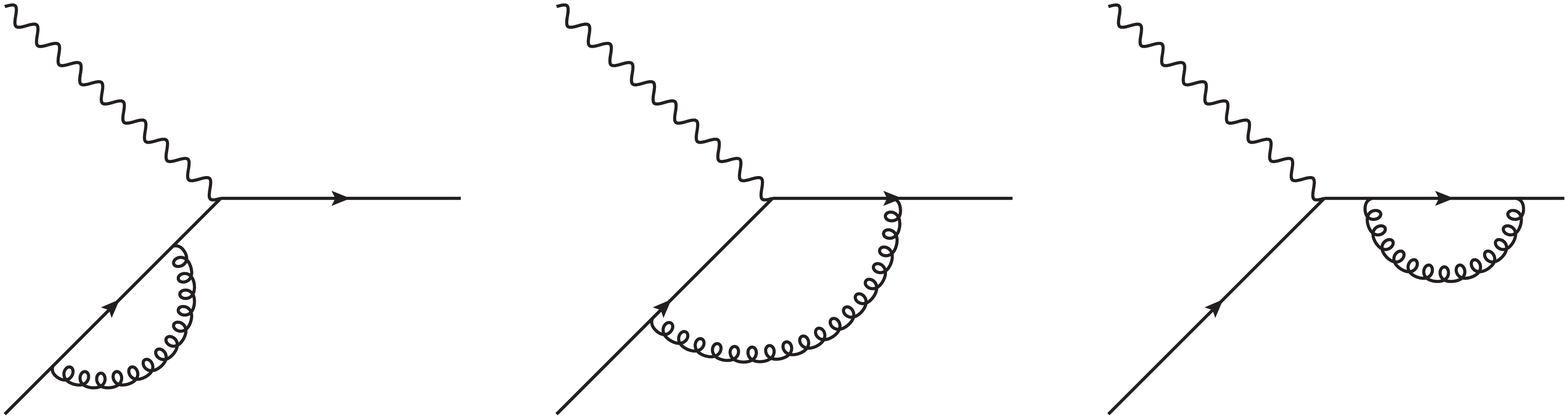}\hspace{1cm}
\end{center}
 \caption{The NLO virtual correction diagrams in SIDIS.}
 \label{fig-virtual}
\end{figure}

We first consider the NLO contribution from the virtual correction which is given by 
the $2\to 1$ scattering amplitude with one gluon loop.
When we adopt the dimensional regularization scheme, the gauge-invariance of the cross section 
is maintained for the loop diagram. Then we can derive the same WTI as shown in Eq. (\ref{WTI}) which is 
the consequence of gauge-invariance of the diagrams. Taking advantage of this fact, we just need to calculate simple diagrams shown in Fig. \ref{fig-virtual} for the virtual correction.
%%%%%%%%%%%%%%%%%%%%%%%%%%%%%%%%%%%%%%%%%%%%%%%%%%%%%%%%%%%%%%%%%%%%%%%%%%%%%%%%%%%%%%%%%%%%%%%%%
%%%%%%%%%%%%%%%%%%%%%%%%%%%%%%%%%%%%%%%%%%%%%%%%%%%%%%%%%%%%%%%%%%%%%%%%%%%%%%%%%%%%%%%%%%%%%%%%% 
The calculation for this kind of one-loop diagrams have been well established, which is exactly the same as the vertex correction at leading twist. We follow the conventional technique here. All ultraviolet divergences can be canceled by the renormalization of the QCD Lagrangian. Then we can set the ultraviolet and infrared divergences are the same with each other in dimensions regularization approach, $\epsilon_{\rm UV}=\epsilon_{\rm IR}$, and identify all divergences 
as infrared. In this definition, we don't have to think about the first and the third amplitudes in Fig. \ref{fig-virtual} because these are exactly zero in the mass case as we considered here. The hard partonic cross section with the second amplitude is given by
\beq
&&\Bigl(-{1\over 1-\epsilon}g_{\mu\nu}\Bigr)
C_Fg^2\mu^{2\epsilon}\int{d^D\ell\over (2\pi)^Di}
{\rm Tr}[x\slash{p}\gamma^{\nu}{\slash{p}_c}\gamma^{\rho}
({\slash{p}_c}-\slash{\ell})\gamma^{\mu}(x\slash{p}-\slash{\ell})
\gamma_{\rho}]{1\over \ell^2({p_c}-\ell)^2(xp-\ell)^2}
\nonumber\\
&=&C_Fg^2\mu^{2\epsilon}\int{d^D\ell\over (2\pi)^Di}\left[
-4{1\over ({p_c}-\ell)^2(xp-\ell)^2}\left({3\over 2}+\epsilon\right)Q^2
+{4Q^2\over \ell^2(p_c-\ell)^2(xp-\ell)^2}\right],
\eeq
where $\epsilon={4-D\over 2}$ in $D$-dimension, we made a change $g_{\mu\nu}\to {1\over 1-\epsilon}g_{\mu\nu}$ for $D$-dimensional calculation, and we used the fact that 
\beq
&&\int{d^D\ell\over (2\pi)^Di}{1\over \ell^2}=\int{d^D\ell\over (2\pi)^Di}{1\over (xp-\ell)^2}
=\int{d^D\ell\over (2\pi)^Di}{1\over \ell^2({p_c}-\ell)^2}
=\int{d^D\ell\over (2\pi)^Di}{1\over \ell^2(xp-\ell)^2}=0.
\eeq
We perform the basic $D$-dimensional calculation for each integration,
\beq
\int{d^D\ell\over (2\pi)^Di}{1\over ({p_c}-\ell)^2(xp-\ell)^2}
&=&{1\over 16\pi^2}\Bigl({4\pi\over Q^2}\Bigr)^{\epsilon}{1\over \Gamma(1-\epsilon)}
\Gamma(1-\epsilon)\Gamma(\epsilon)B(1-\epsilon,1-\epsilon)
\nonumber\\
&=&{1\over 16\pi^2}\Bigl({4\pi\over Q^2}\Bigr)^{\epsilon}{1\over \Gamma(1-\epsilon)}
\Bigl({1\over \epsilon}+2+O(\epsilon)\Bigr),
\eeq

\beq
\int{d^D\ell\over (2\pi)^Di}{1\over \ell^2({p_c}-\ell)^2(xp-\ell)^2}
&=&-{1\over 16\pi^2Q^2}\Bigl({4\pi\over Q^2}\Bigr)^{\epsilon}{1\over \Gamma(1-\epsilon)}
\Gamma(1-\epsilon)\Gamma(1+\epsilon)B(1,-\epsilon)
B(-\epsilon,1-\epsilon)
\nonumber\\
&=&-{1\over 16\pi^2Q^2}\Bigl({4\pi\over Q^2}\Bigr)^{\epsilon}{1\over \Gamma(1-\epsilon)}
\Bigl({1\over \epsilon^2}+O(\epsilon)\Bigr).
\eeq
The complex-conjugate diagram gives the same contribution. 
Then we can show the cross section for the NLO virtual correction as
\beq
{d\la P_{h\perp}\Delta\sigma\ra^{\rm virtual}\over dx_Bdydz_h}
=\tilde\sigma_0 {\alpha_s\over 2\pi}\sum_qe_q^2G_{q,F}(x_B,x_B)D_{q\to h}(z_h)
C_F\Bigr({4\pi\mu^2\over Q^2}\Bigr)^{\epsilon}{1\over \Gamma(1-\epsilon)}
\Bigl(-{2\over \epsilon^2}-{3\over \epsilon}-8\Bigr),
\label{NLOvirtual}
\eeq
this is exactly the same as the virtual correction at leading twist.
The strategy in the virtual correction calculation presented here is different with that shown in Ref. \cite{Kang:2012ns}, in which the authors didn't use the WTI shown in Eq. (\ref{WTI}) and directly calculated $H^{\mu\nu}_{L(R)\rho\,ji}(k_1,k_2)$ 
which has one more external gluon line with the momentum $k_2-k_1$ in Fig. \ref{fig-virtual}.
The authors obtained a consistent result with Eq. (\ref{NLOvirtual}), which demonstrated the validation of WTI through explicit calculation by including all virtual diagrams shown in Figs. 3, 4 in Ref. \cite{Kang:2012ns}. 
We would like to comment that the WTI reduces much calculational cost. The direct calculation of 
$H^{\mu\nu}_{L(R)\rho\,ji}(k_1,k_2)$ takes tremendous time as they contain significant amount of tensor reduction and integration. These two calculations should be conceptually the same with each other as long as we correctly keep track of all imaginary contributions. We confirmed in this section the consistency mathematically in $2\to 1$-scattering case.
The consistency check in a more general way will be a future task in the collinear twist-3 factorization approach.

%-------------------   Section 4   --------------

\subsection{Real correction}

We now complete the NLO calculation by adding the real emission contribution represented by $2\to2$ partonic scattering process. The calculation for $2\to2$ scattering diagrams have been well studied in $P_{h\perp}$-unintegrated case. We just have to repeat the same calculation but in $D$-dimension. 
We adopt the conventional technique by separating the propagator into the principle value part and imaginary part \cite{qiu, Kanazawa:2000cx, koike, Koike:2011mb}, 
\bea
{1\over k^2+i\epsilon}\to P{1\over k^2}-i\pi\delta (k^2),
\eea
and we focus on the pole contribution $-i\delta(k^2)$ which is required to generate the phase space for SSA.
The derivation of the cross section for the pole contribution has been well developed so far
based on the diagrammatic method we reviewed in Sec. 2. Here we recall
the result derived in Ref. \cite{koike} as
\beq
w^{\mu\nu}&=&i\int dx_1\int dx_2\,M^{\alpha}_{ij\,F}(x_1,x_2)p^{\beta}{\partial\over \partial k_2^{\alpha}}
\Bigl(H^{{\rm pole}\,\mu\nu}_{Lji\,\beta}(k_1,k_2)+H^{{\rm pole}\,\mu\nu}_{Rji\,\beta}(k_1,k_2)
\Bigr)\Bigr|_{k_i=x_ip},
\label{poleX}
\eeq
where the matrix element $M^{\alpha}_{ij\,F}$ is given by
\beq
M^{\alpha}_{ij\,F}(x_1,x_2)
&=&\int{d\lambda\over 2\pi}\int{d\lambda'\over 2\pi}e^{i\lambda x_1}e^{i\lambda'(x_2-x_1)}
\la pS_{\perp}|\bar{\psi}_j(0)gF^{\alpha n}(\lambda' n)\psi_i(\lambda n)|pS_{\perp}\ra
\nonumber\\
&=&{M_N\over 4}\left[\epsilon^{\alpha pnS_{\perp}}(\slash{p})_{ij}G_{q,F}(x_1,x_2)
+iS^{\alpha}_{\perp}(\gamma_5\slash{p})_{ij}\tilde{G}_{q,F}(x_1,x_2)\right]+\cdots.
\label{F-type}
\eeq
We can construct the gauge-invariant expression (\ref{poleX}) before performing the $p_h$-integration. 
There are three types of pole contributions in SIDIS 
which are respectively known as soft-gluon pole contribution(SGP, $x_1=x_2=x$), 
hard pole contribution(HP, $x_1=x,\ x_2=x_B$ or $x_1=x_B,\ x_2=x$) and another
hard pole contribution(HP2, $x_1=x_B,\ x_2=x_B-x$ or $x_2=x_B,\ x_1=x_B-x$ )~\cite{koike,Yoshida:2016tfh},
the corresponding hard parts are given by
\beq
H^{{\rm pole}\,\mu\nu}_{Lji\,\beta}
&=&H^{{\rm SGP}\,\mu\nu}_{Lji\,\beta}(k_1,k_2)\Bigl\{-i\pi\delta\Bigl[({p_c}-(k_2-k_1))^2\Bigr]\Bigr\}
(2\pi)\delta\Bigl[(k_2+q-{p_c})^2\Bigr]
\nonumber\\
&&+H^{{\rm HP}\,\mu\nu}_{Lji\,\beta}(k_1,k_2)\Bigl\{-i\pi\delta\Bigl[(k_1+q)^2\Bigr]\Bigr\}
(2\pi)\delta\Bigl[(k_2+q-{p_c})^2\Bigr]
\nonumber\\
&&+H^{{\rm HP2}\,\mu\nu}_{Lji\,\beta}(k_1,k_2)\Bigl\{i\pi\delta\Bigl[(k_2+q)^2\Bigr]\Bigr\}
(2\pi)\delta\Bigl[(k_2-k_1+q-{p_c})^2\Bigr],
\nonumber\\
H^{{\rm pole}\,\mu\nu}_{Rji\,\beta}
&=&H^{{\rm SGP}\,\mu\nu}_{Rji\,\beta}(k_1,k_2)\Bigl\{i\pi\delta\Bigl[({p_c}+(k_2-k_1)]^2\Bigr)\Bigr\}
(2\pi)\delta\Bigl[(k_1+q-{p_c})^2\Bigr]
\nonumber\\
&&+H^{{\rm HP}\,\mu\nu}_{Rji\,\beta}(k_1,k_2)\Bigl\{i\pi\delta\Bigl[(k_2+q)^2\Bigr]\Bigr\}
(2\pi)\delta\Bigl[(k_1+q-{p_c})^2\Bigr]
\nonumber\\
&&+H^{{\rm HP2}\,\mu\nu}_{Rji\,\beta}(k_1,k_2)\Bigl\{-i\pi\delta\Bigl[(k_1+q)^2\Bigr]\Bigr\}
(2\pi)\delta\Bigl[(k_1-k_2+q-{p_c})^2\Bigr].
\eeq
%%%%%%%%%%%%%%%%%%%%%%%%%%%%%%%%%%%%%%%%%%%%%%%%%%%%%%%%%%%%%%%%%%%%%%%%%%%%%%%%%%%%%%%%%%%%%%%%%
\begin{figure}[h]
\begin{center}
  \includegraphics[height=4cm,width=12cm]{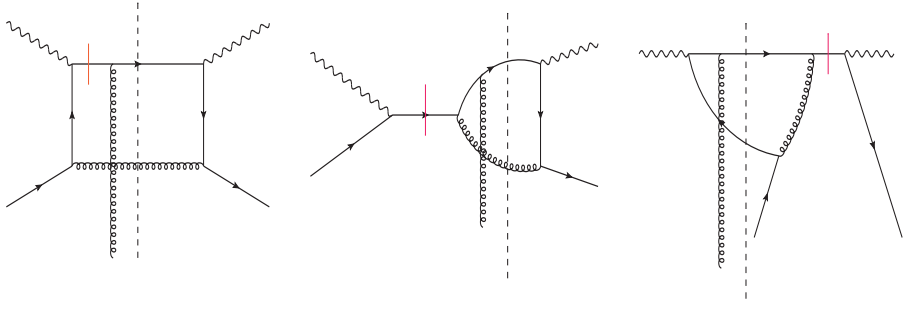}\hspace{1cm}
\end{center}
 \caption{Typical diagrams for soft-gluon pole(left), hard pole(middle) and another hard pole(right).
 The red barred propagator gives the pole term.}
 \label{pole}
\end{figure}
%%%%%%%%%%%%%%%%%%%%%%%%%%%%%%%%%%%%%%%%%%%%%%%%%%%%%%%%%%%%%%%%%%%%%%%%%%%%%%%%%%%%%%%%%%%%%%%%% 
Typical diagrams for each pole contribution are shown in Fig. \ref{pole}
(full diagrams can be found in~\cite{Yoshida:2016tfh}). 
We write down the explicit form of each diagram in Fig. \ref{pole} in order to help readers to 
follow our calculation,
\beq
&&(H^{{\rm SGP}\,\mu\nu}_{Lji\,\beta}(k_1,k_2))_{\rm Fig.\ref{pole}}
\nonumber\\
&=&-[\gamma^{\sigma}(\slash{p}_c-\slash{q})\gamma^{\nu}\slash{p}_c\gamma_{\beta}
(\slash{p}_c-\slash{k}_2+\slash{k}_1)\gamma^{\mu}(\slash{p}_c-\slash{k}_2+\slash{k}_1-\slash{q})
\gamma^{\rho}]_{ji}{1\over (p_c-q)^2}{1\over (p_c-k_2+k_1-q)^2}[-g_{\perp\rho\sigma}(k_2+q-p_c)],
\nonumber\\
&&(H^{{\rm HP}\,\mu\nu}_{Lji\,\beta}(k_1,k_2))_{\rm Fig.\ref{pole}}
\nonumber\\
&=&-[\gamma^{\sigma}(\slash{p}_c-\slash{q})\gamma^{\nu}\slash{p}_c\gamma_{\beta}
(\slash{p}_c-\slash{k}_2+\slash{k}_1)\gamma^{\rho}(\slash{k}_1+\slash{q})
\gamma^{\mu}]_{ji}{1\over (p_c-q)^2}{1\over (p_c-k_2+k_1)^2}[-g_{\perp\rho\sigma}(k_2+q-p_c)],
\nonumber\\
&&(H^{{\rm HP2}\,\mu\nu}_{Lji\,\beta}(k_1,k_2))_{\rm Fig.\ref{pole}}
\nonumber\\
&=&[\gamma^{\nu}(\slash{k}_2+\slash{q})\gamma^{\rho}\slash{p}_c\gamma_{\beta}
(\slash{p}_c-\slash{k}_2+\slash{k}_1)\gamma^{\mu}(\slash{p}_c-\slash{k}_2+\slash{k}_1-\slash{q})
\gamma_{\rho}]_{ji}{1\over (p_c-k_2-q)^2}{1\over (p_c-k_2+k_1)^2},
\eeq
where $g_{\perp\rho\sigma}$ is the sum of the polarization vector
$\sum_r\epsilon_{r\rho}(k)\epsilon_{r\sigma}(k)=-g_{\perp\rho\sigma}(k)$.
We can show the WTI for the pole diagrams
\beq
(k_2-k_1)^{\beta}{H^{{\rm pole}\,\mu\nu}_{L(R)ji\,\beta}(k_1,k_2)}=0,
\eeq
and it gives a useful relation
\beq
p^{\beta}{\partial\over \partial k_2^{\alpha}}H^{{\rm pole}\,\mu\nu}_{L(R)ji\,\beta}(k_1,k_2)
\Bigr|_{k_i=x_ip}=-{1\over x_2-x_1-i\epsilon}H^{{\rm pole}\,\mu\nu}_{L(R)ji\,\alpha}(k_1,k_2).
\label{WTI_pole}
\eeq
For HP and HP2 contributions, we can use Eq. (\ref{WTI_pole}) and don't have to perform 
$k_2$-derivative directly as in Eq. (\ref{poleX}). However, we can't use this relation for SGP contribution
because it contains a delta function $\delta(x_1-x_2)$. In Ref.~\cite{Koike:2006qv}, the authors
found a reduction formula for SGP contribution as
\beq
p^{\beta}{\partial\over \partial k_2^{\alpha}}
\Bigl(H^{{\rm SGP}\,\mu\nu}_{Lji\,\beta}(k_1,k_2)+H^{{\rm SGP}\,\mu\nu}_{Rji\,\beta}(k_1,k_2)\Bigr)
={1\over 2NC_F}{1\over x_2-x_1-i\epsilon}\left({\partial\over \partial p_c^{\alpha}}
-{p_{c\alpha}p^{\mu}\over p_c\cdot p}{\partial\over \partial p_c^{\mu}} \right)H(xp),
\label{master}
\eeq
where $H(xp)$ is the $2\to 2$-scattering cross section without the external gluon line
with momentum $(x_2-x_1)p$.
Substituting Eq. (\ref{F-type}) into Eq. (\ref{poleX}) and using Eqs. (\ref{WTI_pole}, \ref{master}),
we can derive the following result,
\beq
w^{\mu\nu}&=&{M_N\pi^2\over 2}\int{dx\over x}\delta\Bigl[(xp+q-p_c)^2\Bigr]
\Bigl[-2\Bigl((\hat{s}+Q^2)\epsilon^{p_cpnS_{\perp}}+\hat{u}\epsilon^{qpnS_{\perp}}\Bigr){d\over dx}G_{q,F}(x,x){1\over \hat{t}\hat{u}}{\rm Tr}[x\slash{p}H(xp)]
\nonumber\\
&&-2\Bigl[(\hat{s}+Q^2)\epsilon^{p_cpnS_{\perp}}+\hat{u}\epsilon^{qpnS_{\perp}}\Bigr]G_{q,F}(x,x){1\over \hat{t}\hat{u}}\Bigl\{
Q^2\Bigl({\partial\over \partial \hat{s}}-{\partial\over \partial Q^2}\Bigr)
{\rm Tr}[x\slash{p}H(xp)]-{\rm Tr}[x\slash{p}H(xp)]
\Bigr\}
\nonumber\\
&&+G_{q,F}(x,x_B){1\over \hat{x}-1}{\hat{x}\over Q^2}\epsilon_{\alpha}^{\ pnS_{\perp}}
\Bigl[{\rm Tr}[x\slash{p}H_L^{{\rm HP}\,\alpha}(x_Bp,xp)]
+{\rm Tr}[x\slash{p}H_R^{{\rm HP}\,\alpha}(xp,x_Bp)]
\Bigr]
\nonumber\\
&&-\tilde{G}_{q,F}(x,x_B){1\over \hat{x}-1}{\hat{x}\over Q^2}iS_{\perp\alpha}
\Bigl[{\rm Tr}[\gamma_5x\slash{p}H_L^{{\rm HP}\,\alpha}(x_Bp,xp)]
-{\rm Tr}[\gamma_5x\slash{p}H_R^{{\rm HP}\,\alpha}(xp,x_Bp)]\Bigr]
\nonumber\\
&&+G_{q,F}(x_B,x_B-x){\hat{x}\over Q^2}\epsilon_{\alpha}^{\ pnS_{\perp}}
\Bigl[{\rm Tr}[x\slash{p}H_L^{{\rm HP2}\,\alpha}((x_B-x)p,x_Bp)]
+{\rm Tr}[x\slash{p}H_R^{{\rm HP2}\,\alpha}(x_Bp,(x_B-x)p)]
\Bigr]
\nonumber\\
&&-\tilde{G}_{q,F}(x_B,x_B-x){\hat{x}\over Q^2}iS_{\perp\alpha}
\Bigl[{\rm Tr}[\gamma_5x\slash{p}H_L^{{\rm HP2}\,\alpha}((x_B-x)p,x_Bp)]
-{\rm Tr}[\gamma_5x\slash{p}H_R^{{\rm HP2}\,\alpha}(x_Bp,(x_B-x)p)]\Bigr)
\Bigr],
\eeq
where we used the Mandelstam variables
\beq
\hat{s}&=&(xp+q)^2={1-\hat{x}\over \hat{x}}Q^2,
\\
\hat{t}&=&(p_c-q)^2=-{1-\hat{z}\over \hat{x}}Q^2,
\\
\hat{u}&=&(xp-p_c)^2=-{\hat{z}\over \hat{x}}Q^2,
\eeq
with $\hat{x}={x_B\over x}$, $\hat{z}={z_h\over z}$. Then the cross section can be written as 
\beq
&&{d^4\la P_{h\perp}\Delta\sigma\ra^{\rm real}\over dx_Bdydz_h}
\nonumber\\
&\sim&\mu^{2\epsilon}\sum_qe^2_q\int{dz}D_{q\to h}(z)
\int {d^{2-2\epsilon}p_{c\perp}\over (2\pi)^{2-2\epsilon}}
\Bigl[\int {dx\over x}
\delta\left(p^2_{c\perp}-{(1-\hat{x})(1-\hat{z})\hat{z}\over \hat{x}}Q^2\right) 
\nonumber\\
&&\times{1\over 1-\epsilon}\Bigl[{d\over dx}G_{q,F}(x,x)H_D 
+G_{q,F}(x,x)H_{ND}+G_{q,F}(x,x_B)H_{HP}
+\tilde{G}_{q,F}(x,x_B)H_{HPT}
\nonumber\\
&&+G_{q,F}(x_B,x_B-x)H_{HP2}
+\tilde{G}_{q,F}(x_B,x_B-x)H_{HPT2}
\Bigr],
\label{realX}
\eeq
where we used the symmetry of the $P_{h\perp}$-integration as
\beq
\int d^{2-2\epsilon}P_{h\perp}\,P_{h\perp\alpha}P_{h\perp\beta}\epsilon^{\rho\alpha pn}
\epsilon^{\beta pn \sigma}&=&
-\int d^{2-2\epsilon}P_{h\perp}\,{1\over 2(1-\epsilon)}P^2_{h\perp}g_{\perp\alpha\beta}
\epsilon^{\rho\alpha pn}\epsilon^{\beta pn\sigma}.
\label{P_hintegration}
\eeq 
The authors of~\cite{Kang:2012ns} found that the factor $1-\epsilon$ in the denominator is essential to derive 
the correct evolution function of $G_{q,F}(x,x)$. They calculated all SGP and HP contributions 
associated with $G_{q,F}$. After that, 
the HP2 contribution and all $\tilde{G}_{q,F}$ contributions were calculated in~\cite{Yoshida:2016tfh}.
The results of all hard cross sections are listed in~\cite{Yoshida:2016tfh}.
We perform the $p_{c\perp}$-integration,
\beq
&&\int{d^{2-2\epsilon}p_{c\perp}\over (2\pi)^{2-2\epsilon}}\delta
\Bigl[p^2_{c\perp}-{(1-\hat{x})(1-\hat{z})\hat{z}\over \hat{x}}Q^2\Bigr]
\nonumber\\
&=&{1\over (2\pi)^{2-2\epsilon}}\int{dp_{c\perp}}\int d\Omega_{2-2\epsilon}
(p_{c\perp})^{1-2\epsilon}\delta
\Bigl[p^2_{c\perp}-{(1-\hat{x})(1-\hat{z})\hat{z}\over \hat{x}}Q^2\Bigr]
\nonumber\\
&=&{1\over 4\pi}\Bigl({4\pi\over Q^2}\Bigr)^{\epsilon}{1\over \Gamma(1-\epsilon)}
\Bigl[{(1-\hat{x})(1-\hat{z})\hat{z}\over \hat{x}}\Bigr]^{-\epsilon},
\eeq
where $\Omega_{2-2\epsilon}$ is the solid angle and it can be integrated out as
\beq
\int d\Omega_{2-2\epsilon}={2\pi^{1-\epsilon}\over \Gamma(1-\epsilon)}.
\eeq
We carry out the $\epsilon$-expansion as follows.
\beq
\hat{z}^{-\epsilon}&\simeq& 1-\epsilon\ln\hat{z},
\\
\hat{x}^{\epsilon} &\simeq& 1+\epsilon\ln\hat{x},
\\
(1-\hat{z})^{-1-\epsilon}&\simeq& -{1\over \epsilon}\delta(1-\hat{z})+{1\over (1-\hat{z})_+}
-\epsilon\Bigl[{\ln(1-\hat{z})\over 1-\hat{z}}\Bigr]_+,
\\
(1-\hat{x})^{-1-\epsilon}&\simeq& -{1\over \epsilon}\delta(1-\hat{x})+{1\over (1-\hat{x})_+}
-\epsilon\Bigl[{\ln(1-\hat{x})\over 1-\hat{x}}\Bigr]_+.
\eeq
Then the cross section can be derived as follows
\beq
&&{d\la P_{h\perp}\Delta\sigma\ra^{\rm real}\over dx_Bdydz_h}
\nonumber\\
&=&\tilde\sigma_0{\alpha_s\over 2\pi}
\Bigl({4\pi\mu^2\over Q^2}\Bigr)^{\epsilon}{1\over \Gamma(1-\epsilon)}
\sum_qe_q^2\Biggl[C_F{2\over \epsilon^2}G_{q,F}(x_B,x_B)D_{q\to h}(z_h)
\nonumber\\
&&+\Bigl(-{1\over \epsilon}\Bigr)\Biggl\{
D_{q\to h}(z_h)\Bigl\{\int^1_{x_B}{dx\over x}\Bigl[C_F{1+\hat{x}^2\over (1-\hat{x})_+}G_{q,F}(x,x)
+{N\over 2}\Bigl({(1+\hat{x})G_{q,F}(x_B,x)-(1+\hat{x}^2)G_{q,F}(x,x)\over (1-\hat{x})_+}
\nonumber\\
&&+\tilde{G}_{q,F}(x_B,x)\Bigr)\Bigr]
-NG_{q,F}(x_B,x_B)
+{1\over 2N}\int^1_{x_B}{dx\over x}\Bigl((1-2\hat{x})G_{q,F}(x_B,x_B-x)+\tilde{G}_{q,F}(x_B,x_B-x)\Bigr)
\Bigr\}
\nonumber\\
&&+G_{q,F}(x_B,x_B)C_F
\int^1_{z_h}{dz\over z}{1+\hat{z}^2\over (1-\hat{z})_+}D_{q\to h}(z)\Biggr\}+{\rm finite\ terms},
\label{NLOreal}
\eeq
where the derivative term was converted to the nonderivative term through the partial integral,
\beq
\int^{1}_{x_B}dx{d\over dx}G_{q,F}(x,x)(1+\hat{x}^2)=\int^1_{x_B}{dx\over x}G_{q,F}(x,x)
(2\hat{x}^2-2\delta(1-\hat{x})).
\eeq
%From the complete NLO cross section obtained from (\ref{LOcross}), (\ref{NLOvirtual}) and (\ref{NLOreal}), we find that the collinear singularity is completely eliminated under the following renormalization conditions.
Combine the real (Eq. (\ref{NLOvirtual})) and virtual corrections (Eq. (\ref{NLOreal})), we find that the double-pole $1/\epsilon^2$ term, which represents a soft-collinear divergence, cancel out. The remaining collinear divergences, represented by the single pole $1/\epsilon$, can be eliminated by the redefinition of Qiu-Sterman function and fragmentation function, 
\beq
G_{q,F}(x_B,x_B)
&=&G_{q,F}^{(0)}(x_B,x_B)+{\alpha_s\over 2\pi}
\Bigr(-{1\over \hat{\epsilon}}\Bigr)\Bigl\{
\int^1_{x_B}{dx\over x}\Bigl[P_{qq}(\hat{x})G_{q,F}(x,x)
\nonumber\\
&&+{N\over 2}\Bigl({(1+\hat{x})G_{q,F}(x_B,x)-(1+\hat{x}^2)G_{q,F}(x,x)\over (1-\hat{x})_+}
+\tilde{G}_{q,F}(x_B,x)\Bigr)\Bigr]
-NG_{q,F}(x_B,x_B)
\nonumber\\
&&+{1\over 2N}\int^1_{x_B}{dx\over x}\Bigl[(1-2\hat{x})G_{q,F}(x_B,x_B-x)+\tilde{G}_{q,F}(x_B,x_B-x)\Bigr]
\Bigr\},
\\
D_{q\to h}(z_h)&=&D^{(0)}_{q\to h}(z_h)+{\alpha_s\over 2\pi}
\Bigr(-{1\over \hat{\epsilon}}\Bigr)\int^1_{z_h}{dz\over z}P_{qq}(\hat{z})D_{q\to h}(z),
\eeq
where $P_{qq}(x)$ is well known $q\to q$ splitting function 
\beq
P_{qq}(x)=C_F\Bigl[{1+x^2\over (1-x)_+}+{3\over 2}\delta(1-x)\Bigr],
\eeq
and we adopted the $\overline{\rm MS}$-scheme
\beq
{1\over \hat{\epsilon}}={1\over \epsilon}-\gamma_E+\ln4\pi.
\eeq
This collinear singularity is consistent with that of $G_{q,F}(x,x)$ explored 
in \cite{Kang:2008ey,Zhou:2008mz,Braun:2009mi,Schafer:2012ra,Ma:2012xn,Kang:2012em}.
After the renormalization, the final result for the transverse-momentum-weighted SSAs at NLO is given by
\beq
&&{d\la P_{h\perp}\Delta\sigma\ra^{\rm LO+NLO}\over dx_Bdydz_h}
\nonumber\\
&=&\tilde\sigma_0\sum_qe_q^2
\Biggl[G_{q,F}(x_B,x_B,\mu)D_{q\to h}(z_h,\mu)
+{\alpha_s\over 2\pi}\ln\Bigl({Q^2\over \mu^2}\Bigr)\Biggl\{D_{q\to h}(z_h,\mu)\Bigl\{
\int^1_{x_B}{dx\over x}\Bigl[P_{qq}(\hat{x})G_{q,F}(x,x,\mu)
\nonumber\\
&&+{N\over 2}\Bigl({(1+\hat{x})G_{q,F}(x_B,x,\mu)-(1+\hat{x}^2)G_{q,F}F(x,x,\mu)\over (1-\hat{x})_+}
+\tilde{G}_{q,F}(x_B,x,\mu)\Bigr)\Bigr]
\nonumber\\
&&-NG_{q,F}(x_B,x_B,\mu)
+{1\over 2N}\int^1_{x_B}{dx\over x}\Bigl((1-2\hat{x})G_{q,F}(x_B,x_B-x,\mu)
\nonumber\\
&&+\tilde{G}_{q,F}(x_B,x_B-x,\mu)\Bigr)
\Bigr\}+G_{q,F}(x_B,x_B,\mu)
\int^1_{z_h}{dz\over z}P_{qq}(\hat{z})D_{q\to h}(z,\mu)\Biggr\}
\nonumber\\
&&+{\alpha_s\over 2\pi}\int^1_{x_B}{dx\over x}\int^1_{z_h}{dz\over z}
\Biggl\{x{dx\over x}G_{q,F}(x,x,\mu)D_{q\to h}(z,\mu)
{1\over 2N\hat{z}}\Bigl[1-\hat{z}+{(1-\hat{x})^2+2\hat{x}\hat{z}\over (1-\hat{z})_+}
\nonumber\\
&&-\delta(1-\hat{z})\Bigl((1+\hat{x}^2)\ln{\hat{x}\over 1-\hat{x}}+2\hat{x}\Bigr)\Bigr]
+G_{q,F}(x,x,\mu)D_{q\to h}(z,\mu){1\over 2N\hat{z}}\Bigl[
-2\delta(1-\hat{x})\delta(1-\hat{z})
\nonumber\\
&&+{2\hat{x}^3-3\hat{x}^2-1\over (1-\hat{x})_+(1-\hat{z})_+}
+{1+\hat{z}\over (1-\hat{x})_+}-2(1-\hat{x})
+\delta(1-\hat{z})\Bigl(
-(1-\hat{x})(1+2\hat{x})\log{\hat{x}\over 1-\hat{x}}
\nonumber\\
&&-2\Bigl({\ln(1-\hat{x})\over 1-\hat{x}}\Bigr)_+
+{2\over (1-\hat{x})_+}
-2(1-\hat{x})
+2{\ln\hat{x}\over (1-\hat{x})_+}
\Bigr)
+\delta(1-\hat{x})\Bigl(
(1+\hat{z})\ln\hat{z}(1-\hat{z})
\nonumber\\
&&-2{\ln\hat{z}\over (1-\hat{z})_+}
-2\Bigl({\ln(1-\hat{z})\over 1-\hat{z}}\Bigr)_+
+{2\hat{z}\over (1-\hat{z})_+}\Bigr)
\Bigr]
\nonumber\\
&&+G_{q,F}(x,x_B,\mu)D_{q\to h}(z,\mu)\Bigl(C_F+{1\over 2N\hat{z}}\Bigr)\Bigl[
2\delta(1-\hat{x})\delta(1-\hat{z})
+{1+\hat{x}\hat{z}^2\over (1-\hat{x})_+(1-\hat{z})_+}
\nonumber\\
&&+\delta(1-\hat{z})\Bigl(
\log{\hat{x}\over 1-\hat{x}}+2\Bigl({\ln(1-\hat{x})\over 1-\hat{x}}\Bigr)_+
-2{\ln\hat{x}\over (1-\hat{x})_+}-{1+\hat{x}\over (1-\hat{x})_+}
\Bigr)
\nonumber\\
&&+\delta(1-\hat{x})\Bigl(
-(1+\hat{z})\ln\hat{z}(1-\hat{z})+2\Bigl({\ln(1-\hat{z})\over 1-\hat{z}}\Bigr)_+
+2{\ln\hat{z}\over (1-\hat{z})_+}-{2\hat{z}\over (1-\hat{z})_+}
\Bigr)
\Bigr]
\nonumber\\
&&+\tilde{G}_{q,F}(x,x_B,\mu)D_{q\to h}(z,\mu)\Bigl(C_F+{1\over 2N\hat{z}}\Bigr)\Bigl[
-{1-\hat{x}\hat{z}^2\over (1-\hat{x})_+(1-\hat{z})_+}
+\delta(1-\hat{z})\Bigl(
\ln{\hat{x}\over 1-\hat{x}}+3
\Bigr)
\Bigr]
\nonumber\\
&&+G_{q,F}(x_B,x_B-x,\mu)D_{q\to h}(z,\mu)\Bigl[{1\over 2N\hat{z}}\Bigl(
{(1-2\hat{x})\hat{z}^2\over (1-\hat{z})_+}
-\delta(1-\hat{z})(1-2\hat{x})
(\ln{\hat{x}\over 1-\hat{x}}+1)\Bigr)
\nonumber\\
&&+{1\over 2\hat{z}}(1-2\hat{x})\{(1-\hat{z})^2+\hat{z}^2\}
\Bigr]+\tilde{G}_{q,F}(x_B,x_B-x,\mu)D_{q\to h}(z,\mu)\Bigl[{1\over 2N\hat{z}}\Bigl(
{\hat{z}^2\over (1-\hat{z})_+}
\nonumber\\
&&-\delta(1-\hat{z})(\ln{\hat{x}\over 1-\hat{x}}+3)\Bigr)
-{1\over 2\hat{z}}(1-2\hat{x})
\Bigr]-8C_F\delta(1-\hat{x})\delta(1-\hat{z})\Biggr\}
\Biggr]+O(\alpha^2_s).
\eeq
Note that the cross section above doesn't include the contribution from the gluon fragmentation channel.
From the requirement that the physical cross section doesn't depend on the factorization scale
$\mu$,
\beq
{\partial\over \partial\ln \mu^2}
{d\la P_{h\perp}\Delta\sigma\ra^{\rm LO+NLO}\over dx_Bdydz_h}=0,
\label{artificial}
\eeq
we can derive the LO evolution equation for $G_{q,F}(x,x)$,
\beq
&&{\partial\over \partial\ln \mu^2}G_{q,F}(x_B,x_B,\mu^2)
={\alpha_s\over 2\pi}\Bigl\{
\int^1_{x_B}{dx\over x}\Bigl[P_{qq}(\hat{x})G_{q,F}(x,x,\mu^2)
\nonumber\\
&&+{N\over 2}\Bigl({(1+\hat{x})G_{q,F}(x_B,x,\mu^2)-(1+\hat{x}^2)G_{q,F}(x,x,\mu^2)\over (1-\hat{x})_+}
+\tilde{G}_{q,F}(x_B,x,\mu^2)\Bigr)\Bigr]
-NG_{q,F}(x_B,x_B,\mu^2)
\nonumber\\
&&+{1\over 2N}\int^1_{x_B}{dx\over x}\Bigl((1-2\hat{x})G_{q,F}(x_B,x_B-x,\mu^2)
+\tilde{G}_{q,F}(x_B,x_B-x,\mu^2)\Bigr)
\Bigr\}.
\label{complete}
\eeq
This evolution equation based on the transverse-momentum-weighted technique was first discussed in Drell-Yan process~\cite{Vogelsang:2009pj}, in which the authors succeeded in deriving the $G_{q,F}$ terms in the parenthesis $[\cdots]$ in the evolution equation. After that, the authors of~\cite{Kang:2012ns} pointed out that an extra term $-NG_{q,F}(x_B,x_B,\mu^2)$ also contribute to the evolution equation. The HP2 pole contribution and all $\tilde{G}_{q,F}$ terms were obtained in \cite{Yoshida:2016tfh} within the method of transverse momentum weighting.

%-------------------- Section 5 -----------------

\section{Application to other processes}
\label{sec-app}

A lot of works on the transverse-momentum-weighted SSA have been done in recent years. We briefly summarize all related work in this section.
The evolution equation of the Qiu-Sterman function Eq. (\ref{complete}) is still missing the the gluon mixing contribution 
associated with the 3-gluon distribution functions defined by
\beq
&&\int{d\lambda\over 2\pi}\int{d\lambda'\over 2\pi}e^{i\lambda x_1}e^{i\lambda'(x_2-x_1)}
\la pS_{\perp}|d_{bca}F_b^{\beta n}(0)gF_c^{\gamma n}(\lambda' n)F_a^{\alpha n}(\lambda n)|pS_{\perp}\ra
\nonumber\\
&=&2M_N\Bigl[O(x_1,x_2)g^{\alpha\beta}\epsilon^{\gamma pnS_{\perp}}
+O(x_2,x_2-x_1)g^{\beta\gamma}\epsilon^{\alpha pnS_{\perp}}
+O(x_1,x_1-x_2)g^{\alpha\gamma}\epsilon^{\beta pnS_{\perp}}\Bigr]+\cdots,\hspace{5mm}
\\
&&\int{d\lambda\over 2\pi}\int{d\lambda'\over 2\pi}e^{i\lambda x_1}e^{i\lambda'(x_2-x_1)}
\la pS_{\perp}|if_{bca}F_b^{\beta n}(0)gF_c^{\gamma n}(\lambda' n)F_a^{\alpha n}(\lambda n)|pS_{\perp}\ra
\nonumber\\
&=&2M_N\Bigl[N(x_1,x_2)g^{\alpha\beta}\epsilon^{\gamma pnS_{\perp}}
-N(x_2,x_2-x_1)g^{\beta\gamma}\epsilon^{\alpha pnS_{\perp}}
-N(x_1,x_1-x_2)g^{\alpha\gamma}\epsilon^{\beta pnS_{\perp}}\Bigr]+\cdots,
\eeq
where $d_{bca}$ and $if_{bca}$ are the structure constants of $SU(N)$ group.
Fig. 4 shows typical diagrams which give the gluon mixing contribution in SIDIS and Drell-Yan.
%%%%%%%%%%%%%%%%%%%%%%%%%%%%%%%%%%%%%%%%%%%%%%%%%%%%%%%%%%%%%%%%%%%%%%%%%%%%%%%%%%%%%%%%%%%%%%%%%
\begin{figure}[h]
\begin{center}
  \includegraphics[height=4cm,width=12cm]{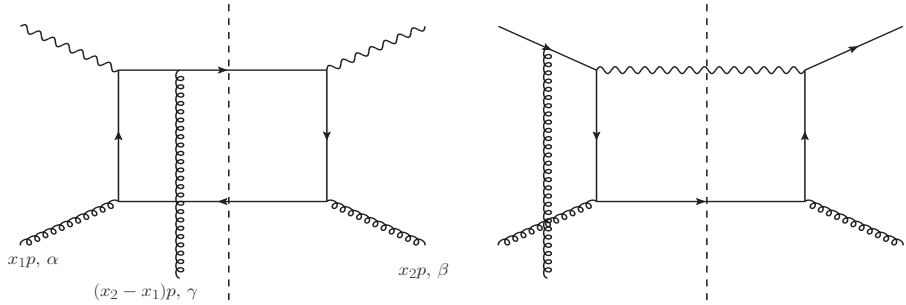}\hspace{1cm}
\end{center}
 \caption{The diagrams which give the gluon mixing contribution to the evolution equation
 of $G_{q,F}(x,x)$ in SIDIS(left) and Drell-Yan(right).}
\end{figure}
%%%%%%%%%%%%%%%%%%%%%%%%%%%%%%%%%%%%%%%%%%%%%%%%%%%%%%%%%%%%%%%%%%%%%%%%%%%%%%%%%%%%%%%%%%%%%%%%% 
The calculation technique for these diagrams was developed in \cite{Koike:2011mb}.
There is only the soft gluon pole($x_1=x_2$) contribution due to the interchange symmetry of 
the external gluon lines and then the cross section is expressed by four independent functions
$O(x,x)$, $O(x,0)$, $N(x,x)$ and $N(x,0)$.
The gluon mixing term in the flavor singlet evolution was discussed 
in both SIDIS~\cite{Dai:2014ala,Chen:2017lvx} and Drell-Yan~\cite{Chen:2016dnp}
and the cross section up to the finite term is derived as
\beq
{d\la P_{h\perp}\Delta\sigma\ra^{\rm SIDIS}\over dx_Bdydz_hd\phi}\Bigr|_{\rm gluon}
&\sim&G_{q,F}(x_B,x_B,\mu)D_{q\to h}(z_h,\mu)+{\alpha_s\over 2\pi}\ln\Bigl({Q^2\over \mu^2}\Bigr)
\Bigl[D_{q\to h}(z_h,\mu){\cal F}_g\otimes T_{G+}\Bigr]+{\rm finite\ terms},
\nonumber\\
{d\la q_{\perp}\Delta\sigma\ra^{\rm DY}\over dydQ^2}\Bigr|_{\rm gluon}
&\sim&G_{q,F}(x_a,x_a,\mu)f_{\bar{q}/p}(x_b,\mu)+{\alpha_s\over 2\pi}\ln\Bigl({Q^2\over \mu^2}\Bigr)
\Bigl[f_{\bar{q}/p}(x_a,\mu){\cal F}_g\otimes T_{G+}\Bigl]+{\rm finite\ terms},
\eeq
where $x_a={Q\over \sqrt{s}}e^{\eta}$, $x_b={Q\over \sqrt{s}}e^{-\eta}$ with the center of mass 
energy $\sqrt{s}$ and the rapidity $\eta$, $T_{G+}$ is given by a linear combination of the 3-gluon distribution functions and 
$f_{\bar{q}/p}$ is the antiquark PDF. In Drell-Yan process, we use the transverse momentum of the virtual
photon $q_{\perp}$ for the weighted cross section. Using the condition as in Eq. (\ref{artificial}), we can derive 
the gluon mixing term of $G_{q,F}$. The explicit form of the evolution kernel ${\cal F}_g$ and 
the finite terms are shown in~Refs. \cite{Chen:2016dnp,Chen:2017lvx}.
Adding the mixing term to Eq. (\ref{complete}), the evolution equation for the Qiu-Sterman function, the first moment of the TMD Sivers function, was completed at LO with respect to QCD coupling constant $\alpha_s$. As a by-product 
of the work on Drell-Yan, the NLO cross section related to the first momentum of 
the TMD Boer-Mulders function was also derived as
\beq
{d\la q_{\perp}\Delta\sigma\ra^{\rm DY}\over dydQ^2}\Bigr|_{\rm BM}
&\sim&T_{q,F}^{(\rho)}(x_a,x_a,\mu)h_1^{\bar{q}}(x_b,\mu)+{\alpha_s\over 2\pi}\ln\Bigl({Q^2\over \mu^2}\Bigr)
\Bigl[h_1^{\bar{q}}(x_a,\mu){\cal F}_{\sigma}\otimes T_{q,F}^{(\rho)}\Bigl]+{\rm finite\ terms},
\eeq
where $h_1$ is chiral-odd transversity distribution and the twist-3 function $T_F^{(\rho)}(x,x,\mu)$ corresponds to the first moment of the 
Boer-Mulders function and its exact definition can be found in~Eq. \cite{Chen:2016dnp}.
The evolution kernel ${\cal F}_{\sigma}$ and 
the finite terms of the cross section are shown in~Ref. \cite{Chen:2016dnp}.
The authors in~Ref. \cite{Chen:2017lvx} also discussed the evolution equation for the 
first moment of the TMD Collins fragmentation function. The calculation of 
the twist-3 fragmentation contribution
is different from the distribution case because the cross section only receives the non-pole 
contribution of the hard scattering. The calculation technique for
 the non-pole contribution has been well developed 
in~Ref. \cite{Yuan:2009dw}. It's not difficult to extend it to the 
$D$-dimensional calculation. The NLO cross section was derived in~Ref. \cite{Chen:2017lvx} as
\beq
{d\la P_{h\perp}\Delta\sigma\ra^{\rm SIDIS}\over dx_Bdydz_hd\phi}\Bigr|_{\rm Collins}
&\sim&h_1^q(x_B,\mu)\hat{e}_{\partial}^q(z_h,\mu)+{\alpha\over 2\pi}\ln\Bigl({Q^2\over \mu^2}\Bigr)
h_1^q(x_B,\mu)\Bigl[{\cal F}_{\partial}\otimes \hat{e}^q_{\partial}
+{\cal F}_{F}\otimes \hat{E}^q_F+{\cal F}_G\otimes \hat{E}^q_G\Bigr]
\nonumber\\
&&+{\rm finite\ terms},
\eeq
where $\hat{e}_{\partial}$, $\hat{E}$ and $\hat{E}_G$ are the twist-3 fragmentation functions
for spin-$0$ hadron(different definition is shown in~Ref. \cite{Kanazawa:2015ajw}).
All evolution kernels and finite terms are shown in~Ref. \cite{Chen:2017lvx}.

The NLO transverse-momentum-weighted cross section has been completed for single inclusive 
hadron production in SIDIS and Drell-Yan process in proton-proton collisions by a series of work
presented here. These results are useful not only for the derivation of the evolution equations,
but also for the verification of twist-3 collinear factorization feasibility, as well as for the global analysis of the experimental data. Measurement of the weighted SSAs 
just began very recent~\cite{Alexeev:2018zvl} and more data will be provided 
by future experiments. The analysis of the data based on the NLO result will 
lead to better understanding of the origin of the SSAs. 
There are still some TMD functions which have not been discussed yet, we hope the techniques presented in this paper can help extending the application of the transverse-momentum-weighted technique to resolve all these open questions.

The transverse-momentum-weighted technique has also been extended to study the transverse momentum broadening effect for semi-inclusive hadron production in lepton-nucleus scattering \cite{Kang:2014ela} and Drell-Yan dilepton production in proton-nucleus scattering \cite{Kang:2016ron}. In these studies, the QCD evolution equation for twist-4 quark-gluon correlation function was derived for the first time, and the twist-4 (double scattering) collinear factorization at NLO was confirmed through explicit calculations \cite{Kang:2013raa}. The finite NLO hard parts were obtained for the transverse momentum broadening effect, which can be used in global analysis of world data to extract precisely the medium properties characterized by the twist-4 matrix elements.

%--------------------- Summary ----------------------

\section{Summary}
\label{sec-sum}

We reviewed the transverse-momentum-weighted technique as a useful tool to derive 
the scale evolution equation for the twist-3 collinear function which is expressed by the 
first moment of the TMD function. We first demonstrated the calculation of the LO cross section
formula in a pedagogical way. Then we showed the basic techniques for the NLO calculation for 
both the virtual correction and real emission contributions. 
A lot of work have been done on the Qiu-Sterman function \cite{Vogelsang:2009pj,Kang:2012ns,Yoshida:2016tfh,Dai:2014ala} and recently 
the application of this technique to other twist-3 functions was also discussed~\cite{Chen:2016dnp,Chen:2017lvx}. 
There is still room of the application to many other TMD functions by considering appropriate twist-3 observables.
We hope our review paper will provide basic knowledge needed to work on this subject.

In the end, we would like to point out the importance of the $P_{h\perp}$-weighted
SSA from the phenomenological point of view. We introduced this observable 
as a tool to derive the scale evolution equation by focusing on the ${1\over \epsilon}$-term
in the cross section. 
However, the finite term is also important when we evaluate the cross section in order to 
compare it with the experimental data. The COMPASS experiment reported the data of the $P_{h\perp}$-weighted SSA
very recent~\cite{Alexeev:2018zvl}. We expect that the data will be accumulated in the future experiments at COMPASS, JLab and EIC 
and then the exact NLO cross section including the finite contribution will play an important role in the analysis 
of those weighted SSA data.

%--------------------- Acknowledgment ----------

\section*{Acknowledgments}

This research is supported by NSFC of China under Project No. 11435004 and research startup funding at SCNU.

\end{document}